\documentclass{article}

\usepackage{arxiv}

\usepackage[utf8]{inputenc} 
\usepackage[T1]{fontenc}    
\usepackage{hyperref}       
\usepackage{url}            
\usepackage{booktabs}       
\usepackage{amsfonts}       
\usepackage{nicefrac}       
\usepackage{microtype}      
\usepackage{lipsum}

\usepackage{hyperref,graphicx,float,booktabs,array,tabularx,multirow,color}
\usepackage{caption,amsmath,siunitx,mathtools,soul, lineno}

\title{Characterization of a Silicon Drift Detector for High-Resolution Electron Spectroscopy}

\author{
  Matteo Gugiatti\\
  Dipartimento di Elettronica, Informazione e Bioingegneria,  Politecnico di Milano, 20133 Milano, Italy\\
  INFN, Sezione di Milano, 20133 Milano, Italy\\
  \texttt{matteo.gugiatti@polimi.it}\\
   \And
 Matteo Biassoni\\
  Dipartimento di Fisica, Università di Milano-Bicocca, 20126 Milano, Italy\\
  INFN, Sezione di Milano-Bicocca, 20126 Milano, Italy\\
  \And
 Marco Carminati\\
  Dipartimento di Elettronica, Informazione e Bioingegneria, Politecnico di Milano, 20133 Milano, Italy\\
  INFN, Sezione di Milano, 20133 Milano, Italy\\
  \texttt{marco1.carminati@polimi.it}\\
  \And
 Oliviero Cremonesi\\
  Dipartimento di Fisica, Università di Milano-Bicocca, 20126 Milano, Italy\\
  INFN, Sezione di Milano-Bicocca, 20126 Milano, Italy\\
  \And
 Carlo Fiorini\\
  Dipartimento di Elettronica, Informazione e Bioingegneria, Politecnico di Milano, 20133 Milano, Italy\\
  INFN, Sezione di Milano, 20133 Milano, Italy\\
  \And
 Pietro King\\
  Dipartimento di Elettronica, Informazione e Bioingegneria\\
  Politecnico di Milano, 20133 Milano, Italy\\
  INFN, Sezione di Milano, 20133 Milano, Italy\\
  \And
 Peter Lechner\\
  Halbleiterlabor of the Max-Planck Society\\
  Munchen, 81739, Germany\\
  \And
 Susanne Mertens\\
  Max Planck Institute for Physics,\\
  80805 Munchen, Germany\\
  Technische Universitat Munchen\\
  80333 Munchen, Germany\\
  \And
 Lorenzo Pagnanini\\
  Dipartimento di Fisica\\
  Università di Milano-Bicocca, 20126 Milano, Italy\\
  INFN, Sezione di Milano-Bicocca, 20126 Milano, Italy\\
   \And
 Maura Pavan\\
  Dipartimento di Fisica\\
  Università di Milano-Bicocca, 20126 Milano, Italy\\
  INFN, Sezione di Milano-Bicocca, 20126 Milano, Italy\\
   \And
 Stefano Pozzi\\
  Dipartimento di Fisica\\
  Università di Milano-Bicocca, 20126 Milano, Italy\\
  INFN, Sezione di Milano-Bicocca, 20126 Milano, Italy\\
}

\begin{document}
\maketitle

\newpage

\begin{abstract}
Silicon Drift Detectors, widely employed in high-resolution and high-rate X-ray applications, are considered here with interest also for electron detection. The accurate measurement of the tritium beta decay is the core of the TRISTAN (TRitium Investigation on STerile to Active Neutrino mixing) project. This work presents the characterization of a single-pixel SDD detector with a mono-energetic electron beam obtained from a Scanning Electron Microscope. The suitability of the SDD to detect electrons, in the energy range spanning from few keV to tens of keV, is demonstrated. Experimental measurements reveal a strong effect of the detector's entrance window structure on the observed energy response. A detailed detector model is therefore necessary to reconstruct the spectrum of an unknown beta-decay source.
\end{abstract}

\section{Introduction}
\label{sec:introduction}

The ability to precisely measure the energy spectra of electrons, originating from radioactive isotopes, would open new frontiers for the $\beta$-decay spectroscopy. The accurate energy reconstruction of $\beta$-decay spectra has a remarkable impact in the field of nuclear physics, in the field of neutrino physics, and in the double $\beta$-decay investigation.

A detection system for electron spectroscopy, with both high-resolution and high-rate capabilities, has wide applications. Among all, the one leading in the research presented in this paper is the TRISTAN project \cite{altenmuller2018silicon}, where the tritium $\beta$-decay spectrum is measured, to search for a keV-scale sterile-neutrino signature \cite{mertens2015sensitivity}. For this application, since a post-acceleration is applied to shift at higher energies the Tritium spectrum, the energy range of interest starts from few \si{keV} up to \SI{30}{keV}, the targeted energy resolution is $<$ \SI{300}{eV} FWHM @ \SI{30}{keV}, and the average count rate is \SI{100}{kcps} per channel \cite{brunst2019development,mertens2019novel}. Other detector requirements, such as radiation hardness for operation in the focal plane of the experiment, are still under definition.

High-resolution and high-count-rate capabilities are characteristic features of the well consolidated Silicon Drift Detector (SDD) technology \cite{lechner2001silicon}, widely employed to precisely resolve X-ray lines \cite{bertuccio2016x,guazzoni2010first,quaglia2015silicon}. Combining a large area coverage and small anode capacitance, these fast detectors are widely adopted for high-resolution X-ray spectroscopy, for photons with energy comprised between few hundreds of eV up to 20-\SI{30}{keV} \cite{hafizh2019characterization}. This paper aims at proving the potentialities of the Silicon Drift Detector, extended in the field of electron detection, for precision $\beta$-decay spectroscopy.


While studies of SDD detection for low-energy X-rays have been carried out \cite{lechner1995ionization,hartmann1996low}, in literature, a limited number of articles report the study of the Silicon Drift Detector response, and other Si detectors \cite{wustling2006large}, to electrons. The response to Internal Conversion Electrons (ICEs) has been measured, using $^{137}$Cs and $^{131\text{m}}$Xe radioactive sources \cite{zhang2020application}, with energies ranging from 129 to \SI{656}{keV}. A different research reports the response to other ICEs originating from $^{109}$Cd, $^{133}$Ba, and $^{133}$Xe sources \cite{cox2011electron}, exploring an additional energy range down to \SI{45}{keV}. However, to our knowledge, a systematic study of the SDD aiming at building a model of the detector response to electrons is still missing.

This work presents a methodical characterization of a single pixel SDD performed by using an artificial electron source, the measurements have been conducted with discrete energy settings between 5 and \SI{20}{keV}. In this energy range, the electronic noise and the incomplete energy absorption, due to dead layer effects, have the most relevant effect on the measured spectra. Spanning from the optimization of the biasing voltages of the detector, to maximise its detection efficiency, to high-statistics measurements of mono-energetic electrons with various energies and incidence angles, a comprehensive set of experimental data is built and here reported. This data is the starting point to build and validate, by means of Geant4 Monte Carlo simulations, a precise physics model of the detector. This model will allow to faithfully reconstruct an unknown $\beta$-decay spectrum starting from its experimentally measured data. The ensemble of detector, characterization method, and detector model, constitutes the basis for the Tritium $\beta$-decay measurement in the TRISTAN project, and it is of great interest also in other applications such as particle and nuclear physics \cite{biassoni2019novel}.

The paper is organised as follows. In Section 2, the experimental setup employed in this work will be reported, from the hardware to the operation point of view. In the following Section 3, multiple sets of experimental measurements are shown: optimisation of the SDD's voltages, rise time of the signal, and detector response to various $e^{\text{-}}$ incidence angles. In the last section 4, a preliminary analytical entrance window model of the SDD is described. Finally, some conclusions are drawn.

\section{Characterization Setup}
\label{sec:setup}
A mono-energetic $e^{\text{-}}$ beam is a suitable source to characterize the response of the SDD to electrons. In contrast, $\beta$-decay isotopes emit electrons with a broad energy range and are hence less suitable for the study of the detector’s response. In alternative, isotopes which emit ICEs can be viable candidates as mono-energetic electron sources, \textit{e.g.} $^{83\text{m}}$Kr does emit electrons at \SI{18}{keV} and \SI{30}{keV} which is in our energy range of interest.

In this work, we use electrons created in an electron scanning microscope (SEM). The electrons are created via the thermionic effect and are then accelerated to a well defined energy through an electric field. The electron beam is subsequently deflected by a system of electromagnetic coils to scan the portion of the sample under analysis. Finally, by the detection of the electrons backscattered by the sample, the electron image is point-by-point reconstructed.
A SEM, by combining a variable electron rate and energy, and a collimated and steerable beam, is the ideal characterization source for low-energy electrons.

\subsection{Detection System}
\label{sec:det_setup}
\begin{figure}[h]
\centering{}
\includegraphics[scale=1.33]{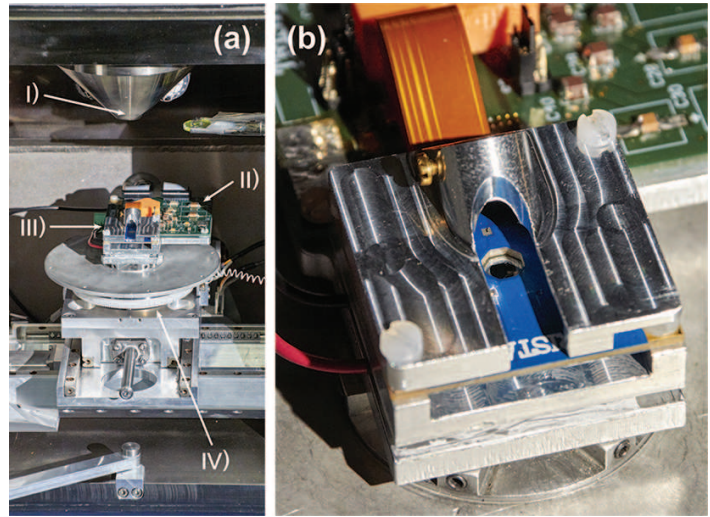}
\caption{(a) Experimental setup installed in the 
SEM. The system is composed of: I) electron beam source, II) board hosting the ASIC charge preamplifier and the filter capacitors, III) detection module with the single SDD, IV) moveable sample holder.
(b) Close up view of the board hosting the SDD, seen from the entrance window side, and the aluminium cover to protect the detector and to host an $^{55}$Fe calibration source.}
\label{fig:setup} 
\end{figure}

A dedicated setup has been developed for our SEM, model Tescan VEGA TS 5136XM, and a picture of its installation is shown in Fig. \ref{fig:setup}. It consists of the electron source, the preamplifier board, and the detection module with a single-pixel SDD. The preamplifier board is hosting the filter capacitors and the Ettore ASIC \cite{trigilio2018ettore}, an integrated charge preamplifier designed to readout the SDD with integrated JFET for the TRISTAN project.

The detection module and the preamplifier board are fixed on a single aluminium frame which is mounted on the sample holder of the microscope. The sample holder allows to move the detector assembly along its x, y, and z axes, including tilt (with respect to the $e^{\text{-}}$ beam) and rotation.
The connections towards the preamplifier board are made with flex cables to allow the free movement of the stage. The cables end on a custom designed 27-pin vacuum feedthrough on the inner wall of the chamber, responsible to transfer all the signals and bias voltages to the outside electronics.

The detector board is hosting a wedge-bonded single-pixel SDD with a diameter of \SI{3.2}{mm} operated in pulsed reset regime. The board is a special-made rigid-flex with three layers of flexible circuit permitting an easy and reliable connection to the preamplifier board, and offering a good signal integrity against crosstalk. Above the detector board, a machined aluminium cover is fixed in place with nylon screws. The cover has a protection purpose and embeds a slot for an $^{55}$Fe source, which provides uncollimated photons for calibration purposes. The cover is made of a conductive material and it is connected to ground in order to avoid, during the time of the measurement, the accumulation of electrons on its surface, a phenomenon which has been observed using a plastic material. Below the SDD board, there is a second aluminium support which is thermally contacting a Peltier cell for cooling the detector. In this paper, however, the measurements were taken at room temperature, obtaining a sufficiently good energy resolution of \SI{190}{eV} FWHM @\SI{5.9}{keV}, for X-rays, with \SI{1}{\micro s} filter peaking time. The removal of the heat generated by the thermoelectric cell, on a moving stage and inside the vacuum environment of the SEM, is critical and needs to be addressed with a custom-made liquid cooling strategy. 

\begin{figure}[ht]
\centering{}
\includegraphics[scale=1.12]{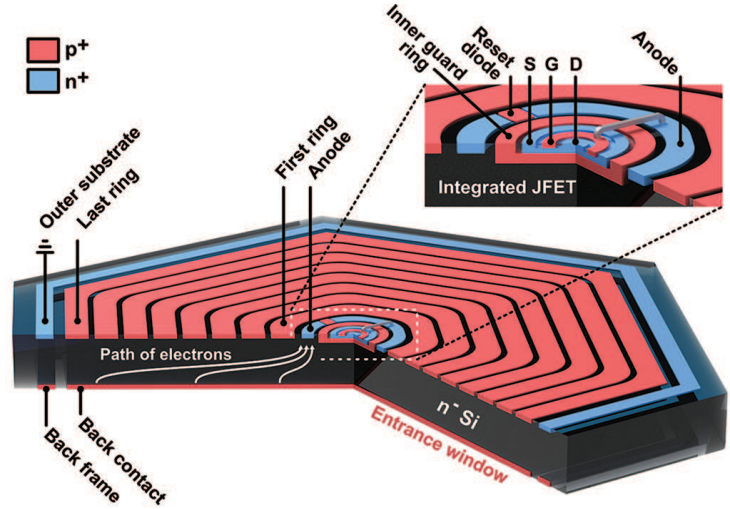}
\caption{Structure of the SDD with integrated JFET, and its contacts, employed in the TRISTAN project.}
\label{fig:SDD_model} 
\end{figure}

The detector employed in the measurements is fabricated by the Semiconductor laboratory of the Max-Planck Society (MPG-HLL). Its entrance window is a thin implanted diode covered by a \SI{22}{nm}-thick SiO$_{2}$ insulating layer without any other additional layer. The integrated JFET is biased by a drain current I$_{\text{D}}\,$=$\,$\SI{300}{\micro A} and has a nominal transconductance g$_{\text{m}}\,$=$\,$\SI{300}{\micro S}. The illustration in Fig. \ref{fig:SDD_model} represents the structure of the silicon device describing the name of its contacts and the doping of its regions: p$^{+}$ corresponds to \SI{1.45e20}{cm^{-3}}. The wafer is a high resistivity one (\SI{6}{k\Omega \cdot cm}) with standard thickness (\SI{450}{\micro m}).

\begin{figure}[ht]
\centering{}
\includegraphics[scale=0.55]{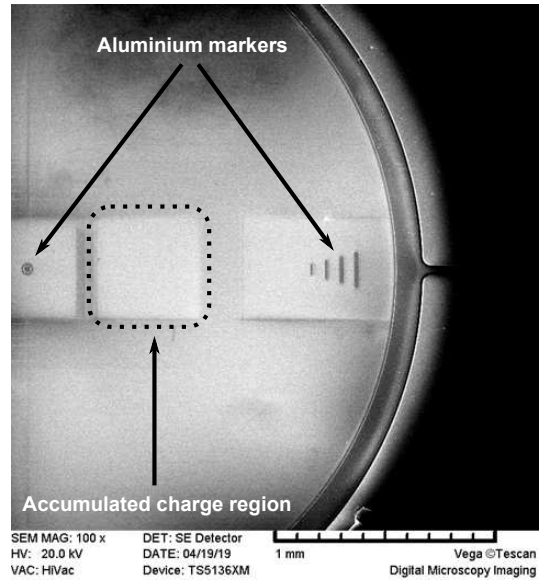}
\caption{Image of the SDD's entrance window acquired by the SEM. The aluminium features, present on the SDD's surface, have been used as a position reference in the measurements. Light rectangular shapes, caused by surface charge accumulation, are visible.}
\label{fig:SDD_surface} 
\end{figure}

A portion of the SDD's entrance window, imaged by the SEM, is visible in Fig. \ref{fig:SDD_surface}. Aluminium markers are deposited on the entrance window, during the manufacturing phase, and are used as an accurate reference for the $e^{\text{-}}$ beam position.
In the same figure an interesting effect can be observed: the presence of lighter rectangular shapes on the uniform SDD's entrance window. This effect is likely due to the temporary accumulation of charges (electrons) in the passivation layer on the SDD. Trapping of excess charge in SiO$_{2}$ under SEM illumination is a well known effect and depends on several parameters of the layer such as permittivity, density of defects and stress \cite{gong1993charge}. However, no differences in the spectra have been observed from measurements inside and outside these slightly charged regions. The effect is reversible and fades away, with a time constant in the order of some minutes, if the entrance window is left unexposed. Overall, the effect is not considered to be an issue for two reasons: first, it is reversible and does not create permanent damage in the short term, and second, when the SEM is used as an $e^{\text{-}}$ source for characterization, the beam current is lowered by several orders of magnitude with respect to its nominal \SI{40}{\micro A} value used for imaging purposes. For comparison, a \SI{40}{\micro A} $e^{\text{-}}$ beam would lead to a rate of \SI{250e12}{cps}, much higher than the \SI{100e3}{cps/pixel} rate required by the TRISTAN project.
During imaging, an area of \SI{400}{\micro m} $x$ \SI{400}{\micro m} is scanned in 3 minutes by the beam focused on a \SI{1}{\micro m} spot, corresponding to a charge of about \SI{45}{nC} impinging on each spot for \SI{1.1}{ms} (comparable to what reported in \cite{gong1993charge}). In order to achieve a charge of \SI{0.45}{nC/\micro m^{2}} (well below a critical threshold for the dielectric), during the operation as focal plane detector of the TRISTAN experiment, an exposure time in excess of 7000 years would be required.

\subsection{External Electronics}
\label{sec:extelectr}

The setup described so far, in section \ref{sec:det_setup}, is contained in the microscope's chamber. Outside, the electronic chain is completed by power supplies, the bias system and the DAQ. The bias system provides all the required voltages to supply the ASIC preamplifier and the SDD. The manual adjustment of the following detector voltages is allowed: V$_\text{BC}$ (back contact voltage), V$_\text{BF}$ (back frame voltage), V$_\text{R1}$ ($1^\text{st}$ ring voltage), V$_\text{RX}$ (last ring voltage), V$_\text{IGR}$ (inner guard ring voltage), V$_\text{D}$ (JFET drain voltage), V$_\text{H}$ (SDD reset diode high level), and V$_\text{L}$ (SDD reset diode low level). The bias system also includes an amplifying stage for the signal G$\,$=$\,$+3  before the DAQ system. The electronics is powered by two bench-top power supplies: $\pm10\,$V for the low voltages, and $+150\,$V from which are derived all the SDD's high voltages. A signal generator is employed to select the reset period of the SDD and the charge preamplifier, to adapt to various rate and leakage current conditions. The DAQ is a commercial single-channel DPP (Digital Pulse Processor), DANTE by XGLab, implementing a trapezoidal shaping filter.

\subsection{Operation of the SEM}

During the normal imaging mode of a SEM, the electron beam is scanned across the area to be imaged using an $e^{\text{-}}$ current in the order of tens of \si{\micro A} (\SI{40}{\micro A} for our specific model). This operation mode is not suitable for our scope, where a very low-current and position-fixed beam is required. In order to obtain an $e^{\text{-}}$ beam with those characteristics, the manual adjustment of the microscope's parameters is needed. First, to reduce the count rate, the $e^{\text{-}}$ current is decreased by reducing the heating power, thus the temperature, of the filament emitting the electrons. Secondly, the beam position is set to fixed coordinates in the point of interest. At this stage, any imaging capability of the SEM is lost, but the beam has acquired the characteristics needed for our use. A collimated beam with a spot size of $\sim$\SI{100}{nm} and an $e^{\text{-}}$ current in the order of \SI{1}{fA} is obtained. This current corresponds to an average rate on the detector of few kcps. The energy of the beam is selectable, changing the acceleration voltage, between the following fixed values: 5, 10, and \SI{20}{keV}.

\section{Experimental Measurements}

This section presents the experimental results of the SDD's characterization with electrons. All the measurements have been carried out inside the microscope's chamber (pressure $<$10$^{-4}$ mbar), at room temperature, with the simultaneous presence of a collimated mono-energetic $e^{\text{-}}$ beam and uncollimated X-rays, with \SI{1}{\micro s} filter peaking time.

\begin{figure}[ht]
\centering{}
\includegraphics[scale=0.2]{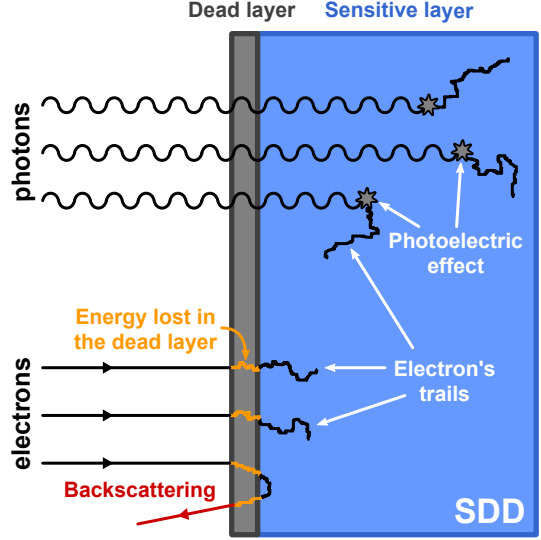}
\caption{Qualitative comparison between the absorption of photons and electrons in a Si detector. The photoelectrons generated by the photons are mostly absorbed inside the device, whereas the electrons arriving from the outside are absorbed in proximity of the entrance window and are subject to energy loss and the backscattering effect.}
\label{fig:phvselectron}
\end{figure}

\begin{figure}[h]
\centering{}
\includegraphics[scale=0.7]{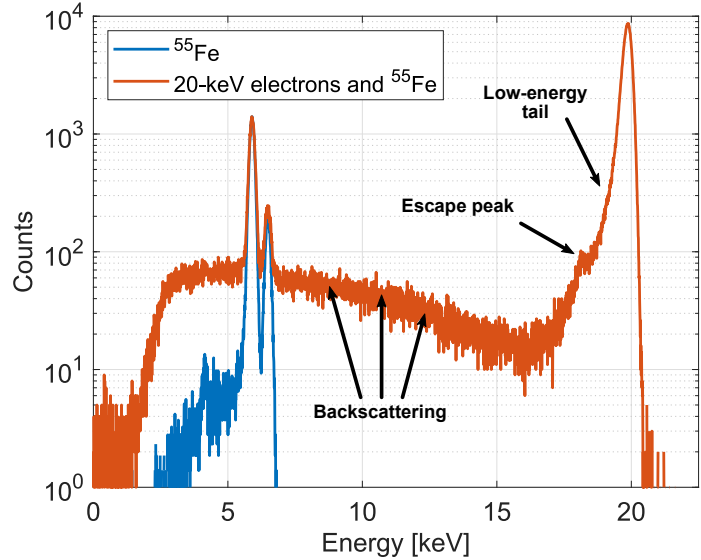}
\caption{Comparison between $^{55}$Fe K$\alpha$ and K$\beta$ X-ray lines and mono-energetic 20-keV electrons having normal incidence with respect to the entrance window surface.}
\label{fig:fevselectronspectra} 
\end{figure}

The different nature of photons and electrons determines the characteristic energy response, to these particles, observed by the SDD. For X-ray photons, the photoelectric absorption mechanism is dominant, while electrons are directly converted to electron-hole pairs along their trail in the detector. Fig. \ref{fig:phvselectron} illustrates the absorption mechanism for the two types of particles. The absorption of a photon generates a photoelectron of equal energy which, for 5.9 keV photons, is statistically well inside in the detector's active volume (for very-low-energy photons the absorption occurs very close to the entrance windows instead). The spectrum originating from a typical X-ray line, neglecting the charge losses during the collection of the photoelectrons, can be considered a Gaussian function. Electrons, instead, have to interact first with the entrance window before releasing their energy into the active volume, where they will, eventually, come at rest. A part of their energy is always released in the superficial layer of the SDD and cannot be measured. Furthermore, there is the possibility to have electrons which release a part of their energy, in the active volume, and then backscatter still retaining a considerable fraction of their initial kinetic energy. The effect of the dead layer and backscattering lead to different features in the response. The dead layer determines the shift, to lower energies, and the asymmetry of the electron peak. Whereas, the backscattering creates a low-energy continuum in the spectrum. The experimental measurement in Fig. \ref{fig:fevselectronspectra} shows the response to X-rays only and to X-rays with a 20-keV $e^{\text{-}}$ beam focused \SI{750}{\micro m} away from the centre of the device. The energy resolution obtained with the setup, at room temperature, is \SI{190}{eV} FWHM at the $^{55}$Fe K$\alpha$ line (\SI{5.9}{keV}) and \SI{265}{eV} FWHM at the 20-keV $e^{\text{-}}$ peak (calculated by fitting the right-hand side of the $e^{\text{-}}$ semi-Gaussian peak).

\subsection{SDD Bias Optimisation}

This section presents a series of measurements aiming at optimizing the detector's biasing voltages to obtain the best performance. The biasing voltages of the SDD shape the electric field inside the detector. The best set of voltages is the one which maximises both the absorption capabilities and the collection of charge carriers generated by the incoming radiation. Two fundamental bias voltages to be optimised are: the V$_\text{BC}$ (back contact) and the V$_\text{RX}$ (last ring) voltages which control the depletion of the detector's volume. The best method to optimise the bias parameters is through a series of measurements where each bias voltage is changed step by step and the peak position, of a reference incoming radiation, is monitored: the voltage range that yields to the highest centroid position is the good operating region where the electric field optimally collects the generated electron-hole pairs.

\begin{figure}[ht]
\centering{}
\includegraphics[scale=0.7]{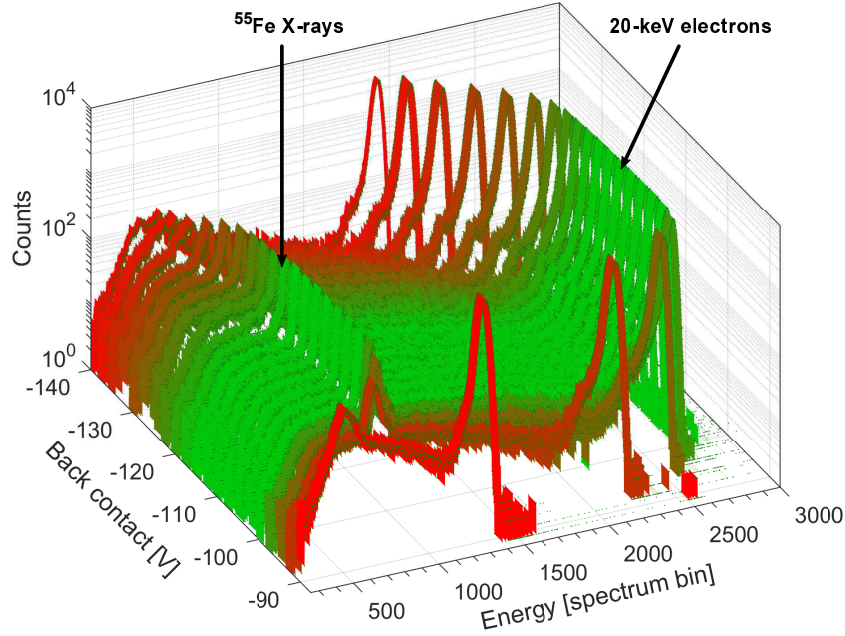}
\caption{Series of experimental spectra with different V$_\text{BC}$ biasing conditions. The voltage of the SDD back contact (entrance window) is varied from \SI{-90}{V} to \SI{-140}{V} by \SI{5}{V} steps. The V$_\text{RX}$ contact is at \SI{-125}{V} while the V$_\text{BF}$ contact is always \SI {10}{V} more negative than the V$_\text{BC}$ voltage. The color of each spectrum represents the position (in bins) of the 20-keV electron peak, green is the full energy.}
\label{fig:BCsweep3D}
\end{figure}

The three-dimensional plot in Fig. \ref{fig:BCsweep3D} reports a series of spectra taken at different V$_\text{BC}$ voltages. A green region can be denoted, where the signals of X-rays and electrons are maximised. This region is the best operating condition for the V$_\text{BC}$ voltage of our specific SDD. 

\begin{figure}[ht]
\centering{}
\includegraphics[scale=0.7]{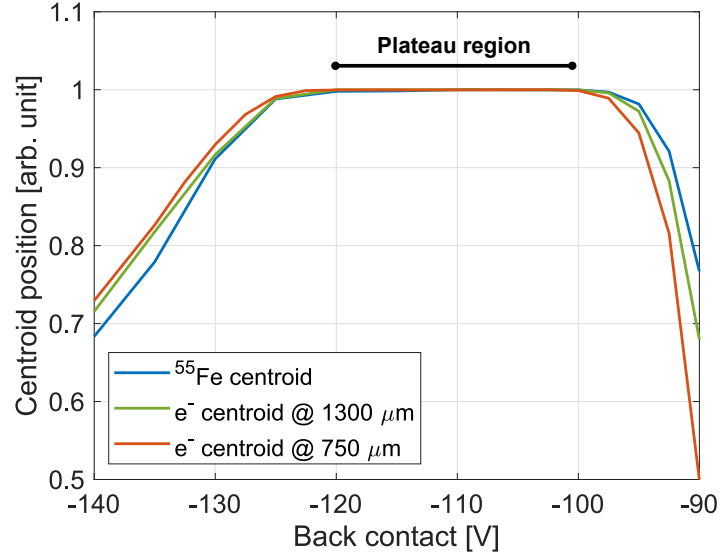}
\caption{Normalised centroid position for calibration X-rays and 20-keV electrons in two positions (middle and outer, respectively \SI{750}{\micro m} and \SI{1300}{\micro m} from the centre of the SDD) versus the V$_\text{BC}$ voltage. V$_\text{RX}=\,$\SI{-125}{V}.}
\label{fig:BCsweep} 
\end{figure}

The V$_\text{BC}$ is necessary to fully deplete the substrate of the SDD and to provide the drifting field towards the anode region, whose potential is kept fixed around \SI{+3}{V}, with respect to ground, by the feedback loop of the preamplifier. Fig. \ref{fig:BCsweep} shows the position of the centroids of the $^{55}$Fe K$\alpha$ and the electron's peak, normalised to their respective maximum values, as a function of the V$_\text{BC}$ voltage. If the V$_\text{BC}$ voltage is less negative ($>$\SI{-100}{V}) some charge starts to be lost due to an insufficient depletion of the device, if V$_\text{BC}$ is too negative ($<$\SI{-125}{V}) a part of the charge cloud is lost in the innermost drift rings. The typical plateau region where the charge collection is optimal can be identified from the plot and the final chosen voltage is V$_\text{BC}=\,$\SI{-110}{V}.

A similar optimisation procedure has been adopted for the SDD's last ring voltage V$_\text{RX}$, keeping V$_\text{BC}=\,$\SI{-110}{V} fixed. The last ring is one of the two terminals of the integrated voltage divider biasing the drift rings. By controlling the V$_\text{RX}$ potential, the radial field in the detector, concentrating the photoelectrons to the anode, changes of magnitude. A minimum voltage in needed to properly drift and collect the charge carriers generated in the whole volume. A maximum voltage limit exists at V$_\text{RX}=-2\cdot |\text{V}_\text{Depletion}|\simeq\,$\SI{-180}{V}, beyond which, a reach-through current arises from the back-contact side of the detector \cite{fiorini2000single}. 

\begin{figure}[ht]
\centering{}
\includegraphics[scale=0.7]{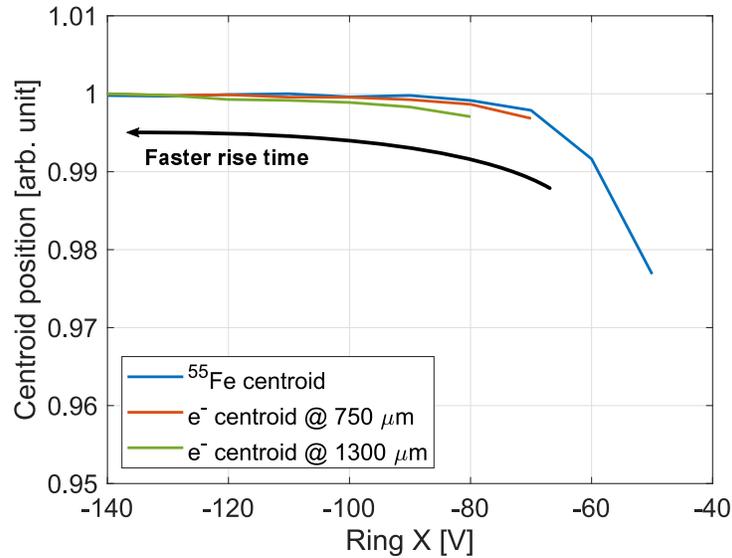}
\caption{Normalised centroid position for calibration X-rays (uncollimated) and 20-keV electrons collimated in two positions (\SI{750}{\micro m} and \SI{1300}{\micro m} from the centre of the SDD) versus the V$_\text{RX}$ voltage. V$_\text{BC}=\,$\SI{-110}{V}.}
\label{fig:RXsweep} 
\end{figure}

In Fig. \ref{fig:RXsweep} the centroid position is shown similarly to the previous measurement, with the V$_\text{RX}$ voltage being a variable parameter. A fraction of uncollimated X-rays is measured, from the central region of the entrance window, at V$_\text{RX}=\,$\SI{-50}{V}. Collimated radiation ($e^{\text{-}}$) \SI{750}{\micro m} from the centre, is successfully measured with V$_\text{RX}=\,$\SI{-70}{V}. Finally, collimated radiation hitting close to the border (\SI{1300}{\micro m}), is properly collected with V$_\text{RX}=\,$\SI{-80}{V}. For V$_\text{RX}<\,$\SI{-80}{V} the device is in optimal working conditions, the whole volume being sensitive, and minimal differences are observed if the bias is further increased. However, an higher radial field is beneficial to decrease the drift time of the charge carries, effect which is confirmed by the measurements presented in section \ref{sec:risetime}. A shorter drift time generates a faster signal which is advantageous because of lower ballistic deficit due to the enlargement of the electron cloud width. Hence, the final value adopted in our setup is V$_\text{RX}=\,$\SI{-140}{V}.

All the remaining SDD's bias voltages, acting in the readout zone (integrated JFET, inner rings, and guard rings), have been optimised in a different laboratory setup to obtain the best $^{55}$Fe energy resolution. In this setup the detector can be cooled down to \SI{-30}{\celsius} reducing the leakage current down to $\sim$\SI{100}{fA}. The complete set of the voltages adopted for our SDD is summarised in Table \ref{tab:SDDbias}. These values are consistent with what is typically obtained from device simulations. The best energy resolution which has been obtained for X-rays, in the cooled laboratory setup, is \SI{127}{eV} FWHM @ \SI{5.9}{keV} with \SI{8}{\micro s} filter peaking time.

\begin{table}[h]
   \centering
   \caption{Complete list of the optimised SDD bias voltages. Please refer to section \ref{sec:extelectr} for the naming of the different voltages.}
   \label{tab:SDDbias}
   \begin{tabular}{cccccccc}
     \hline
     \textbf{V}$_{\text{\textbf{BC}}}$ & \textbf{V}$_{\text{\textbf{BF}}}$ & \textbf{V}$_\text{\textbf{R1}}$ & \textbf{V}$_\text{\textbf{RX}}$ & \textbf{V}$_\text{\textbf{IGR}}$ & \textbf{V}$_\text{\textbf{D}}$ & \textbf{V}$_\text{\textbf{H}}$ & \textbf{V}$_\text{\textbf{L}}$\\
     \hline
     \SI{-110}{V} & \SI{-125}{V} & \SI{-20}{V} & \SI{-140}{V} & \SI{-27.5}{V} & \SI{6}{V} & \SI{4.25}{V} & \SI{-9.5}{V}\\
\end{tabular}
\end{table}

\subsection{Position Measurements}

\begin{figure}[ht]
\centering{}
\includegraphics[scale=0.7]{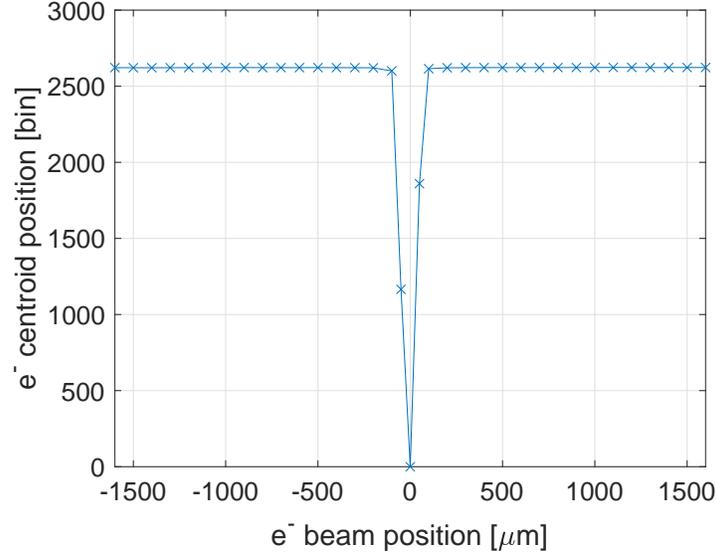}
\caption{Centroid position of the $e^{\text{-}}$ peak versus the position of the beam in different points along the diameter of the SDD. The charge loss in the centre of the detector is visible.}
\label{fig:diametralscan} 
\end{figure}

The availability of a beam with excellent collimation and precise positioning, both in the sub-\si{\micro m} range, allows to study the position-dependence of the detector response. The SDD has been scanned across its 3.2-mm diameter, with a 20-keV $e^{\text{-}}$ beam using \SI{100}{\micro m} steps (\SI{50}{\micro m} steps for the central points), the centroid position has been calculated for each point and the result is plotted as a function of the position. Given the circular shape of the detector, a scan along a single diameter was performed assuming perfect radial symmetry of the device. From the measurements, which are reported in Fig. \ref{fig:diametralscan}, the presence of an insensitive central region of the SDD is clealry visible. The position of the $e^{\text{-}}$ peak, in the spectrum, is rapidly decreasing and disappearing getting closer to the centre of the entrance window. The number of counts drops to zero accordingly.

This behaviour is expected for the SDD used in this measurements, where the charge carriers generated above the anode region are collected by the Drain (the most positive electrode) instead of the anode. From the measurements, the dead spot is a circle with a diameter $<$\SI{200}{\micro m} which is less than $0.5\%$ of the entrance window's total area. Therefore, the impact due to the lost events is negligible. However, in a new production of the detectors for TRISTAN, this effect will be eliminated by an improved design of the integrated read-out structure which prevents the collection of the charges into the JFET electrodes. Outside of the central region, the measured energy of the events is homogeneous and stable over the measurement time of about 30 minutes (all the points are within their statistical oscillations without an observable long-term drift).

\subsection{Rise Time Evaluation}
\label{sec:risetime}

The rise-time performance of the detector connected to the charge preamplifier has been evaluated with a set of dedicated measurements here presented. Instead of using a DPP, the signal is sampled by a very fast digitizer. The detector is illuminated with 20-keV collimated electrons and its output waveform is acquired at \SI{10}{GS/s} with 12-bit depth. The waveforms are then processed in order to extract the rise-time information. The events are detected, from the raw waveform, with a derivative filter, then a selection is made to take only full-energy non-saturating events. Each one is then fitted with a parametric erf (Gaussian error function) and the associated $10$\,-\,$90\%$ rise time is calculated.

\begin{figure}[ht]
\centering{}
\includegraphics[scale=0.7]{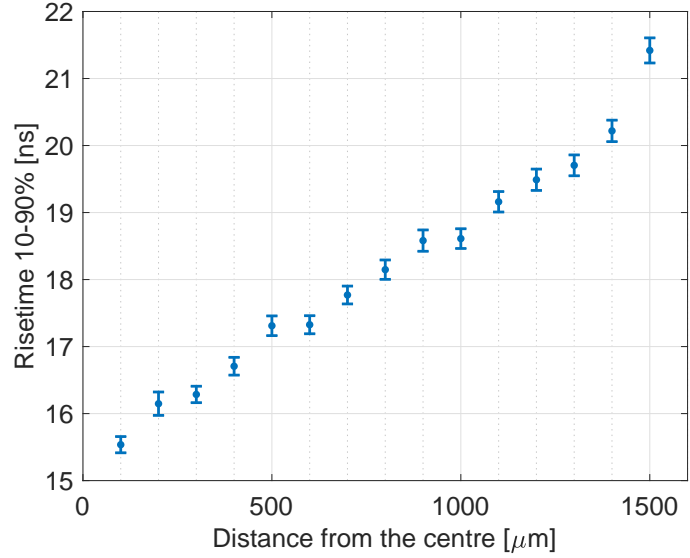}
\caption{Rise time of the signal at the output of the preamplifier for different 20-keV radial beam positions. The error bars indicate the $\pm \sigma$ uncertainties for each point.}
\label{fig:pos_risetime} 
\end{figure}

Fig. \ref{fig:pos_risetime} reports the rise time in different points of irradiation, along the radius of the SDD's entrance window, scanned with the electron beam. The measured rise time is determined by the convolution between the response of the SDD and the transfer function of the electronics. In our case, the electronics is faster (BW $\simeq$ \SI{40}{MHz}) than the detector and the difference in signal width, between particles absorbed at different distances from the anode, can be appreciated. 

The charge cloud, travelling from the $e^{\text{-}}$ interaction point to the anode, guided by the drift field, is subject to a spatial broadening effect which is proportional to the drift time. The charge cloud associated to radiation absorbed far from the anode, requires more time to reach its destination and its broadening is translated into a slower electrical signal \textit{i.e.} slower rise time. The additional information coming from the signal rise time can be, in principle, employed to implement an electronic collimation of the detector. A useful scenario would be the rejection of events which produce charge sharing between adjacent pixels in multi-pixel SDD matrices.

\begin{figure}[ht]
\centering{}
\includegraphics[scale=0.7]{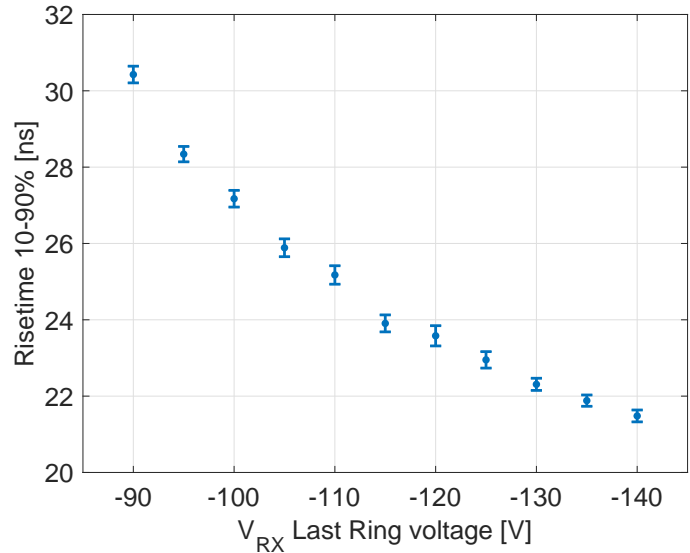}
\caption{Signal rise time versus last ring voltage. The position of the $e^{\text{-}}$ beam is fixed \SI{1500}{\micro m} from the anode. $\pm \sigma$ error bars are represented for each measurement point.}
\label{fig:rx_risetime} 
\end{figure}

Another rise time measurement is reported in Fig. \ref{fig:rx_risetime}, where the rise time is measured as a function of the SDD bias voltage. The electrons are focused on a fixed spot distant \SI{1500}{\micro m} from the centre of the SDD, which is the region of the detector with lower field, so the most sensitive to voltage variations. The last ring voltage is swept from \SI{-90}{V} to \SI{-140}{V} keeping V$_\text{R1}=\,$\SI{-20}{V} and the rise time is measured. For higher magnitudes of the V$_\text{RX}$ biasing voltage, a faster signal is observed.

\subsection{Tilt Measurements}

The setup, mounted on the microscope's sample holder, can be tilted with respect to the $e^{\text{-}}$ beam. The angle can range from zero, when the beam is perpendicular to the entrance window, up to $65 ^{\circ}$ without loosing the line of sight. The amount of energy lost in the passivation layer of the detector, which does not generate any signal, depends on the thickness of the passivation layer itself. If the detector is not perpendicular to the electron beam and it is assumed that electrons travel, in average, in a straight line, the effective insensitive layer that they encounter is increasing with the incidence angle. The backscattering probability of the incoming electrons is also affected by geometrical parameters such as the angle. Both these effects have been experimentally observed.
\\

\begin{figure}[h!]
\centering{}
\includegraphics[scale=0.59]{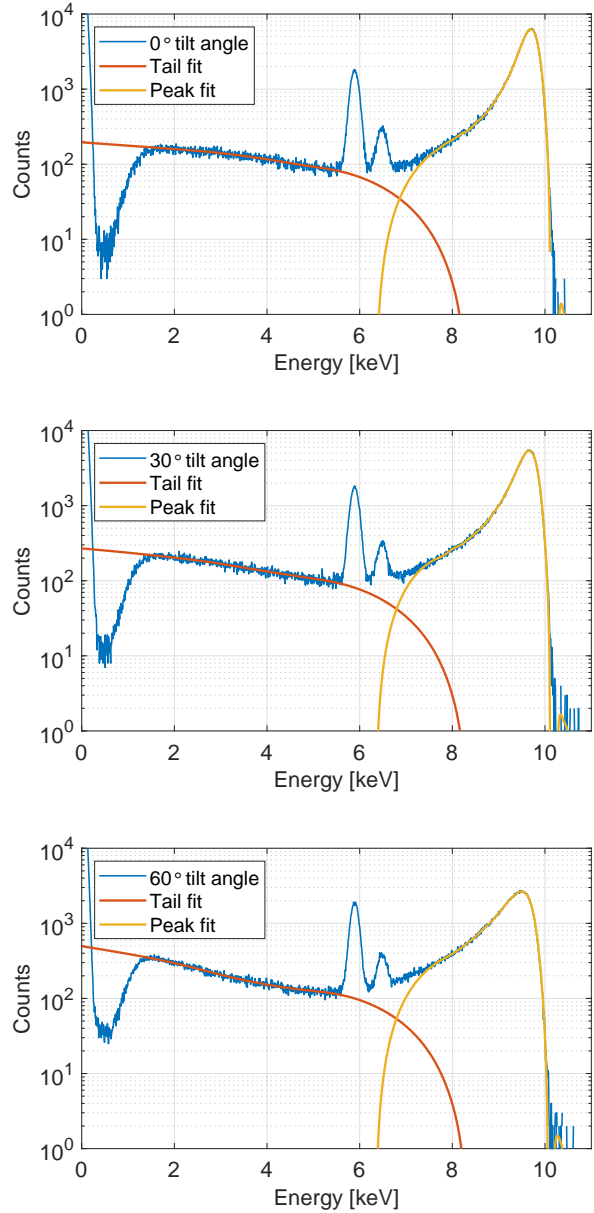}
\caption{Spectra of 10-keV electrons and $^{55}$Fe X-rays for various $e^{\text{-}}$ incidence angles. The energy axis is calibrated on the X-ray lines. The $^{55}$Fe calibration source is fixed in the reference system of the detector.}
\label{fig:TiltSpectra} 
\end{figure}

In Fig. \ref{fig:TiltSpectra} the spectra obtained at $0^{\circ}$, $30^{\circ}$, and $60^{\circ}$ incidence angles, with 10-keV electrons, are plotted in log scale. By increasing the angle of incidence, the energy corresponding to the maximum of the $e^{\text{-}}$ peak is progressively decreasing due to the increased energy loss in the superficial layer of the detector. The fraction of counts in the low-energy continuum is also increasing due to the higher backscattering probability.

\subsubsection{Backscattering}

For each incidence angle, the $e^{\text{-}}$ tail and the peak in the spectra are fitted with cubic spline functions constrained by thresholds defining the end of the tail and the beginning of the peak, while the $^{55}$Fe lines are ignored in the fitting. For energies below \SI{1.5}{keV}, where the DAQ threshold cuts off, the fitting of the tails are interpolated using third-order polynomial functions. For each spectrum, the integral of the fitted tail and the integral of the fitting of the peak are calculated, then the ratio between the tail and the total (tail counts plus peak counts) is determined. 

\begin{figure}[!ht]
\centering{}
\includegraphics[scale=0.7]{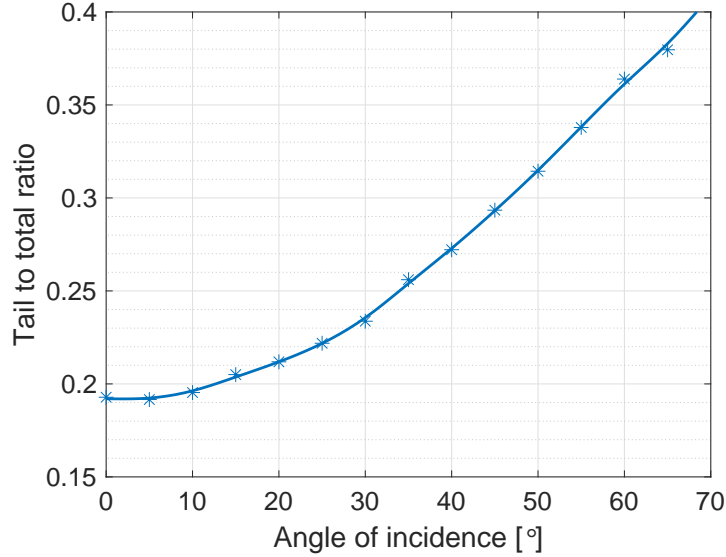}
\caption{Tail to total ratio of the 10-keV electron beam for different incident angles ranging from $0^{\circ}$ to $65^{\circ}$. The fitting of the spectra, in the tail (low-energy continuum) and in the peak, is done as per the algorithm used in Fig. \ref{fig:TiltSpectra}.}
\label{fig:TailToTotal} 
\end{figure}

Fig. \ref{fig:TailToTotal} shows the measured tail-to-total ratio for the electrons as a function of their incidence angle. The observed increasing trend, which is expected, quantitatively shows the higher backscattering probability as the incident angle is increased.

\subsubsection{Energy Loss}
If a simplified Si detector structure is considered, i.e. a thin dead layer with thickness $t$ on top of an ideal sensitive layer, and the propagation of the electrons in the SDD is approximated to a straight line, it is possible to define a geometrical link between the energy loss in the dead layer and the angle of incidence of the electrons. The effective  distance d$_{\text{eff}}$, travelled by $e^{\text{-}}$ in the silicon, as a function of the incidence angle $\alpha$ is geometrically given by (\ref{eq:DeadDistance}), where $\Gamma$ represents the additional factor relative to the dead layer thickness $t$.

\begin{equation}
\ d_{eff} = t (1 + \Gamma) = \dfrac{t}{\cos \alpha} \quad \Rightarrow \quad \Gamma = \frac{1}{\cos \alpha} - 1
\label{eq:DeadDistance} 
\end{equation}

Assuming the energy of the 10-keV electron beam to be accurate within few eV and the potential of the SDD's entrance window being set at V$_\text{BC}=\,$\SI{-110}{V}, with respect to the microscope's ground, the energy of the electrons reaching the detector is E$_{\text{eff}} = \text{E}_{\text{beam}} + \text{V}_{\text{BC}}=\,$\SI{10}{keV} $-$ \SI{0.11}{keV} = \SI{9.89}{keV}. Defining as E$_{\text{meas}}(\alpha)$ the energy of the maximum of the $e^{\text{-}}$ peak, with respect to the X-ray calibration, the energy lost when the beam is perpendicular to the detector is E$_{\text{lost}}(0^{\circ})= \text{E}_{\text{eff}} - \text{E}_{\text{meas}}(0^{\circ})$. Assuming the travel of the electrons being straight, the energy being lost in the dead layer can be described by (\ref{eq:LostEnergy}).

\begin{equation}
E_{lost}(\alpha) = (1 + \Gamma)E_{lost}(0^{\circ})= \frac{1}{\cos \alpha}E_{lost}(0^{\circ})
\label{eq:LostEnergy} 
\end{equation}

In conclusion, the actual energy measured on the electron peak E$_{\text{meas}}(\alpha)$ is described by equation (\ref{eq:MeasEnergy}). With E$_{\text{lost}}(0^{\circ}) =\,$\SI{190}{eV} being determined experimentally from the $\alpha = 0^{\circ}$ data.

\begin{equation}
E_{meas}(\alpha) = E_{eff} - E_{lost}(\alpha) = E_{beam} + V_{BC} - \frac{1}{\cos \alpha}E_{lost}(0^{\circ})
\label{eq:MeasEnergy} 
\end{equation}

\begin{figure}[!ht]
\centering{}
\includegraphics[scale=0.7]{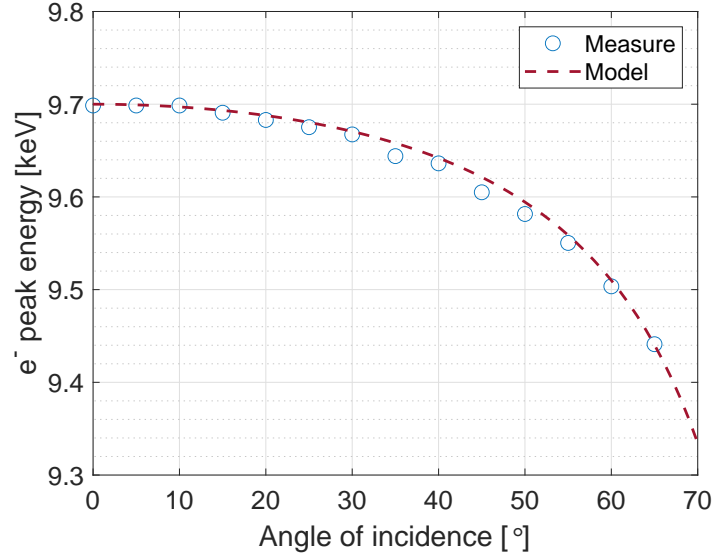}
\caption{Energy of the maximum of the electron peak versus the incidence angle of the 10-keV beam measured with $5^{\circ}$ steps, compared to the model of equation (\ref{eq:MeasEnergy}). The energy axis is calibrated with the $^{55}$Fe X-ray lines.}
\label{fig:MaxEPeak}
\end{figure}

The energy loss changes with the energy of the impinging electrons. Geant4 Monte Carlo simulations allow to accurately model this energy dependence. Here we focus on the effect of the incident angle and take as reference the energy of \SI{10}{keV}.
In Fig. \ref{fig:MaxEPeak} the energy of the maximum of the 10-keV $e^{\text{-}}$ semi-Gaussian peak and the result of the equation (\ref{eq:MeasEnergy}) are plotted as a function of the angle of incidence. A very good agreement is found between the experimental data and the model obtained with the geometrical considerations. The maximum energy residual is $0.14\%$ (\SI{13.3}{eV}) and the root mean square of the residuals is $0.09\%$ (\SI{8.5}{eV}). This confirms the strong impact of the detector's entrance window on the measured energy of the beam and the goodness of the geometrical assumptions made in this section.

\section{Entrance Window Model}
The difficulty in the electron spectroscopy lies in the estimation of the real energy of the detected electrons from the energy effectively measured by the detector. It is demonstrated that the entrance window has a strong effect in the response to electrons, hence a correct modelling of its structure is a critical aspect in this application.

In this work, a comprehensive set of high-quality high-statistics data has been collected in different experimental conditions (\textit{e.g.} various beam energies and incidence angles). This data is used by a closely related work \cite{biassoni2020electron}, which implements and reports a method to model the SDD's entrance window by combining the experimental data with Geant4 \cite{allison2016recent} simulations. The main result, focusing on the detector's structure, is here briefly illustrated.

An analytical function describing the charge collection efficiency (CCE) versus the depth $z$ in the entrance window is defined as per equation (\ref{eq:analytical_model}), which reflects the technology used to build the SDD. The model for the entrance window is known as partial event model, which has been developed for X-ray response \cite{lechner1995ionization, hartmann1996low, popp1999measurement}.

\begin{equation}
    f_\mathrm{CCE}(z;t,p_0,p_1,\lambda) = \begin{dcases*} p_0 & $z < t$ \\ 
        1+(p_1-1)\exp{\left(-\frac{z-t}{\lambda}\right)} &  $z > t$
    \end{dcases*}
    \label{eq:analytical_model} 
\end{equation}

The entrance window is characterised by an oxide layer (SiO$_{2}$) with thickness $t$ = \SI{22}{nm} which is completely insensitive ($p_{0}$ = 0). After the oxide layer, the implantation of the entrance window contact occurs. Here it is assumed a CCE starting from a value $p_{1}$ which gradually approaches the unitary value following an exponential equation with a characteristic constant $\lambda$. $p_{1}$ and $\lambda$ are free parameters in the model.

\begin{figure}[ht]
\centering{}
\includegraphics[scale=0.7]{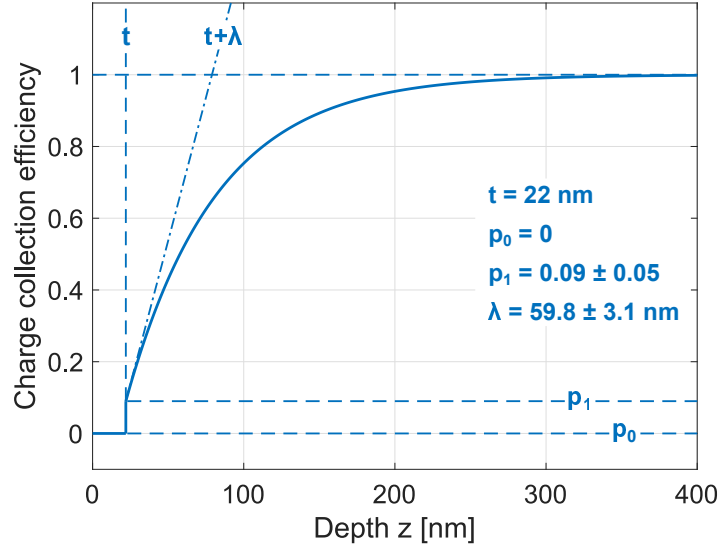}
\caption{Model of the SDD's entrance window describing the charge collection efficiency as a function of the depth in the device. The values of the parameters and their uncertainties are shown.}
\label{fig:window_model}
\end{figure}

Geant4 Montecarlo simulations reproduce a series of spectra spanning all the possible combinations of the free parameters in (\ref{eq:analytical_model}). The values of the free parameters which minimise the discrepancy between the simulated spectra and the experimental spectra are considered to be the optimal parameters to describe the entrance windows model. The model hereby defined can be employed to predict the response of the SDD to any source of electrons. The outcome of this study is shown in Fig. \ref{fig:window_model}. The optimized parameters, and related uncertainties, are $p_{1}$ = 0.09 $\pm$ 0.05 and $\lambda$ = \SI{59.8}{nm} $\pm$ \SI{3.1}{nm}.

\section{Conclusions}

Silicon Drift Detectors despite being considered so far mainly for X-ray detection, can represent an excellent sensor also to measure electrons, with energies ranging from the keV to tens of keV, offering high-count-rate and high-resolution capabilities equivalent to their usual X-ray applications. However, the spectra generated by electrons are considerably different from the ones generated by photons of similar energy. 

The process dominating the electrons energy loss is the ionization along the track. The electron-hole pairs which are created in the most superficial part of the detector cannot be collected by the drift field. A fraction of the energy is always lost in the entrance window and the effect is depending on its structure. Moreover, the non-zero probability of electron backscattering adds a characteristic low-energy continuum due to incomplete energy deposition by the incoming particles. The incidence angle between the electrons and the detector is also playing a role in the shape of the measured spectrum, since the electrons travel through an increased effective thickness of superficial layer. If the $e^{\text{-}}$ beam source is not collimated, it is convenient to place the detector in a space where a combination of electrical and magnetic fields guarantees a good perpendicularity between the charged particles and the SDD's entrance window to avoid the mixing of various incidence angles.

The optimal biasing of an SDD does not change between the measurement of photons or electrons. Once either particle releases its energy, creating $h^{\text{+}}\,$-$\,e^{\text{-}}$ pairs, there are no differences in the charge collection mechanism inside the detector's volume.

Experimental data, acquired during the measurements presented in this work, are the basis to build a model of the SDD detector, which is illustrated, for reconstructing unknown $\beta$-decay spectra. Future developments will include new measurements featuring a multi-pixel SDD matrix to study the effects of charge sharing between adjacent pixels.

\section*{Acknowledgments}

This work has been supported by INFN through TRISTAN and by Max Planck Institute for Physics.
The experimental measurements have been carried out in the Electron Microscopy laboratory inside the Department of Material Science of the University of Milano-Bicocca. We would like to thank Prof. M. Acciarri and Dr. P. Gentile for their competence, support, and patience. 

\bibliographystyle{unsrt}  
\bibliography{references}

\end{document}